\documentclass[preprint,showpacs,amsmath,amssymb,showkeys]{revtex4}

\usepackage{graphicx} % Include figure files
\usepackage{dcolumn} % Align table columns on decimal point
\usepackage{bm} % bold math
\usepackage{color} % allows using colored text

\begin{document}

\preprint{}

\title{Symmetry breaking in a mechanical resonator made from a carbon nanotube}
\author{A. Eichler$^{1,2}$}
\author{J. Moser$^{1,2}$}
\author{M. I. Dykman$^{3}$}
\author{A. Bachtold$^{1,2}$}
\affiliation{$^1$ICFO - Institut de Ciencies Fotoniques,
Mediterranean Technology Park, 08860 Castelldefels, Barcelona,
Spain,} \affiliation{$^2$Institut Catal{\`a} de Nanotecnologia,
Campus de la UAB, E-08193 Bellaterra, Spain,} \affiliation{$^3$
Department of Physics and Astronomy, Michigan State University,
East Lansing, Michigan 48824, USA}
\date{\today}% It is always \today, today, but any date may be explicitly specified

\begin{abstract}
Nanotubes behave as semi-flexible polymers in that they can bend
by a sizeable amount. When integrating a nanotube in a mechanical
resonator, the bending is expected to break the symmetry of the
restoring potential. Here, we report on a new detection method
that allows us to demonstrate such symmetry breaking. The method
probes the motion of the nanotube resonator at (nearly)
zero-frequency; this motion is the low-frequency counterpart of
the second overtone of resonantly excited vibrations. We find that
symmetry breaking leads to the spectral broadening of mechanical
resonances, and to an apparent quality factor $Q$ that drops below
100 at room temperature. This low $Q$ at room temperature is a
striking feature of nanotube resonators whose origin has remained
elusive for many years. Our results shed light on the pivotal role
played by symmetry breaking in the mechanics of nanotube
resonators.

\end{abstract}

\maketitle

% %%%%%%%%%%%%%%%%%%%%%%%%%%%%%%%%%%%%%%%%%%%%%%%%%%%%%%%%%%%%%%%%%%%%%%%%
% \section{Introduction}
% %%%%%%%%%%%%%%%%%%%%%%%%%%%%%%%%%%%%%%%%%%%%%%%%%%%%%%%%%%%%%%%%%%%%%%%%

A carbon nanotube is a unique system that can be seen both as a
crystal and as a polymer. Its crystallinity confers excellent
mechanical properties to nanotube-based
resonators~\cite{Sazonova2004, Lassagne2009,
Steele2009,Gouttenoire2010,Ganzhorn2013,Benyamini}, such as high resonant
frequencies~\cite{Chaste2011, Laird2012} and low dissipation at
low temperature~\cite{Huttel2009, Eichler2011}. As a result, these
resonators are well suited for ultra-sensitive detection of
mass~\cite{Chiu2008, Chaste2012}, charge~\cite{Lassagne2009,
Steele2009} and force~\cite{Moser}. A nanotube has also much in
common with a polymer as both can bend by a large amount. In a
resonator, the bending can be generated by the mechanical tension
that builds in during the fabrication process, as well as by the
electrostatic force used in most studies. This curvature is
expected to have profound consequences on the dynamics of nanotube
resonators, since the transverse vibrational modes lack inversion
symmetry.

In a bent nanotube, if one thinks of a vibrational mode as an
oscillator, its potential is not symmetric with respect to the
displacement from the equilibrium position (Fig.~1a). This leads
to a nonlinear term in the restoring force that depends
quadratically on the displacement, $F_2 = m\beta \cdot \delta
z^2(t)$ with $m$ the effective mass of the resonator, $\beta$ a
constant quantifying the strength of the symmetry breaking effect,
and $\delta z(t)$ the transverse displacement of the resonator for
the given mode. The mechanism underlying the effect can be
understood as follows. For a resonator that is curved and clamped
at both ends, the length is different for $+\delta z$ and $-\delta
z$ and, therefore, the tension induced by the motion is asymmetric
with respect to $\delta z$ (Fig.~1b).

In a potential with broken symmetry, the equilibrium position of
the mode depends on its vibrational amplitude $z_{vibra}$ (see
Fig.~1a). Indeed, if the resonator vibrates as $\delta z(t) =
z_{vibra} \cdot \cos(\omega t)$, the quadratic term in the
restoring force becomes
\begin{align}
\label{eq:force} F_2 =m \beta\frac{z_{vibra}^2}{2}[1 + \cos(2
\omega t)].
\end{align}
The second term in the bracket leads to the second overtone, that
is, motion at $2\omega$. The first term corresponds to a
time-independent force and, therefore, generates a shift of the
equilibrium position, $\delta z_{eq}$. In other words, it is
possible to move the equilibrium position by varying $z_{vibra}$.
This motion is slow, since it is limited by the ring-down time
associated to $z_{vibra}$. When considering the thermal motion of
such a resonator, the power spectrum of the displacement is
expected to feature a peak at zero-frequency, its width being
roughly the inverse of the ring-down time~\cite{Krivoglaz1966,
Dykman1991}. To the best of our knowledge, this low-frequency
motion of the equilibrium position of high-$Q$ oscillators has not
been observed in nanomechanical resonators or other
condensed-matter systems.

The device consists of a single carbon nanotube that is clamped by
two metal electrodes and is suspended over a trench. A gate
electrode is defined at the bottom of the trench (Fig.~1c). The
fabrication is described elsewhere~\cite{Eichler2011}. Briefly, we
pattern the three electrodes and the trench using standard
electron-beam lithography techniques. We grow the nanotube by
chemical vapor deposition (CVD) in the last fabrication step in
order to avoid contamination~\cite{Huttel2009}. All measurements
are performed at $65$\,K to avoid Coulomb
blockade~\cite{Lassagne2009, Steele2009}. We have studied $5$
nanotube devices in total. We discuss in the following the data
for one device. Data for a second device yielding similar results
are shown in supplementary section XI.

We employ a new technique to detect the motion of nanotube
resonators. We capacitively drive the vibrations at
$\omega_{drive}$ near the resonant angular frequency $\omega_{0}$
by applying a constant voltage $V_g^{dc}$ and an oscillating
voltage of amplitude $V_g^{ac}$ on the gate electrode. Central to
the technique is that the oscillating voltage is amplitude
modulated (AM) so that the resulting displacement near
$\omega_{0}$ for not too large $V_g^{ac}$ is proportional to the
driving amplitude,
\begin{align}
\label{eq:motion} \delta z(t) = z_{vibra} \cdot
\cos(\omega_{drive} t-\varphi_m) \cdot [1 - \cos(\omega_{AM} t)]
\end{align}
where the amplitude modulation has a depth of $100\,\%$ and its
angular frequency $\omega_{AM}$ is typically $2\pi\times1$~kHz
(top of Fig.~1d; we checked that the measurements do not depend on
$\omega_{AM}$); $\varphi_m$ is the phase difference between the
displacement and the driving force. We apply a constant voltage
$V_{sd}^{dc}$ to the source electrode and measure from the drain
electrode the low-frequency current $I_{LF}$ at $\omega_{AM}$ with
a lock-in amplifier (see supplementary section VI for details). We
show below that this technique allows us to measure the motion of
$\delta z_{eq}$ associated with the symmetry breaking in nanotube
resonators.

We observe that $I_{LF}$ features a peak when $\omega_{drive}$ is
swept through a mechanical resonance (Fig.~2a); the mechanical
resonance is also verified by directly measuring the vibrational
motion using the frequency modulation (FM) mixing technique
(Fig.~2b)~\cite{Gouttenoire2010}. The height $I_{LF}^{max}$ of the
peak in $I_{LF}$ goes linearly to zero as $V_{sd}^{dc}$ is
decreased (Figs.~2c,d; black squares). These data show that the
detected peak in $I_{LF}$ is related to the modulation of the
nanotube conductance $\delta G = I_{LF}/V_{sd}^{dc}$ at $\omega
_{AM}$. It rules out an artefact related to the capacitive
coupling between the gate and the source electrodes. (This
coupling could result in a sizeable AM oscillating voltage at the
source electrode, and could thus drive the resonator; but $I_{LF}$
would then be independent of $V_{sd}^{dc}$.)

We also observe a current peak when setting the reference
(angular) frequency of the lock-in amplifier to $2\omega_{AM}$.
The peaks measured at $\omega_{AM}$ and $2\omega_{AM}$ are similar
in that they appear at the same driving frequency and their
heights depend linearly on $V_{sd}^{dc}$ (Fig.~2d). However, the
height of the peak measured at $2\omega_{AM}$ is four times
smaller. These observations suggest that the measured peaks are
related to a nonlinearity that scales as ($\delta z(t))^2$ and
therefore $z_{vibra}^2$. (Indeed, if $\delta z(t) \propto 1 -
\cos(\omega_{AM} t)$, any physical quantity that is proportional
to ($\delta z(t))^2$ will be modulated at $\omega_{AM}$ and
$2\omega_{AM}$ with a ratio of $4$ between the amplitudes of the
two components.)

We estimate $z_{vibra} \simeq 2.1$\,nm at resonant frequency for
the driving force used in Fig.~2c. For this, we use the 2-source
mixing technique with a driving voltage of $1.1$\,mV (Fig.~3a,b).
The value of $z_{vibra}$ is inferred by comparing the signals on
and away from resonance~\cite{Eichler2011B} (supplementary section
V).

The peak in $I_{LF}$ is detected only for a fraction of the
mechanical eigenmodes. In the resonator discussed thus far, the
peak is observed for the second eigenmode but not for the first
one (by comparing $I_{LF}$ in Fig.~4a and the current in Fig.~4b
obtained with the FM mixing technique). In all the five studied
resonators, we find that about half of the eigenmodes feature a
peak in $I_{LF}$.

We now discuss different possible origins of the peak in $I_{LF}$.
It could be related to the nonlinear capacitive coupling between
the nanotube and the gate electrode, which leads to a $(\delta
z(t))^2$ nonlinearity in the conductance of the nanotube. However,
we estimate that the current associated to this effect is
$I_{LF}^{capa} = 10$\,pA, which is $20$ times smaller than the
measured value in Fig.~2c (supplementary section VII). Thus, we
reject the $(\delta z(t))^2$ nonlinearity induced by the
capacitive coupling as the physical origin of the peak in
$I_{LF}$. Neither is the $I_{LF}$ peak attributed to the
nonlinearity of the conductance in gate voltage~\cite{Huttel2009},
since it leads to a current that is 3 orders of magnitude lower
than that measured in Fig.~2c (supplementary section VII). Another
mechanism for the $I_{LF}$ peak could be the piezoresistance of
the nanotube, whose dependence on the displacement is quadratic to
a good approximation~\cite{Sansa}. The piezoresistance effect in
nanotubes is by far strongest for positive gate voltages (where
electrons tunnel from the p-doped regions of the nanotube near the
metal electrodes into the n-doped region of the suspended part of
the nanotube~\cite{Minot2003,Stampfer2006}). However, the observed
height of the peak in $I_{LF}$ can be as large for negative as for
positive gate voltages (see Fig.~4a). We can thus rule out the
piezoresistive effect.

We now consider that the peak in $I_{LF}$ is due to symmetry
breaking of the vibrations. For the AM modulation $\propto [1 -
\cos(\omega_{AM} t)]$ the height of the peak in $I_{LF}$ depends
on the maximal displacement of the equilibrium position $\delta
z_{eq}^{0}$ as
\begin{align}
I_{LF}^{max} = V_{sd}^{dc} V_g^{dc}\partial_{V_g} G \cdot
\frac{\partial_z C_g}{C_g} \delta z_{eq}^{0}. \label{AM_current}
\end{align}
where $\delta z_{eq}^{0}$ is proportional to $z_{vibra}^2$. Here,
$\partial_z C_{g}$ is the derivative of the nanotube-gate
capacitance $C_{g}$ with respect to displacement; it is determined
from Coulomb blockade measurements at helium temperature (see
supplementary section III). From the measured $I_{LF}^{max}$ in
Fig.~2c, we get that $\delta z_{eq}^{0}=0.18$\,nm and $\beta= 4.3
\cdot 10^{24}$\,m$^{-1}$s$^{-2}$ using the relation $\beta= \omega
_0^2 \delta z_{eq}^{0}/ z_{vibra}^2$ (see supplementary section
VIII).

This value for the symmetry breaking strength can be compared to
the one estimated from the measurement of $\omega_0$ as a function
of the oscillating driving force. Figures~4c and d show that the
peak in $I_{LF}$ shifts to lower frequency upon increasing the
driving force. Disregarding the cubic restoring force (which in
nanotubes leads to the shift in the opposite direction, see
Ref.~\cite{Eichler2012}), we obtain from the shift in $\omega_0$
that $\beta = 4.1 \cdot 10^{24}$\,m$^{-1}$s$^{-2}$ (supplementary
section VIII). This value agrees with the one estimated from
$I_{LF}^{max}$, demonstrating that the peak in $I_{LF}$ is due to
symmetry breaking of the vibrations.

The strength of symmetry breaking can be made large in nanotubes,
since it scales as $\beta \simeq \frac{E}{\rho}z_s\left(
\frac{\pi}{L}\right)^4$ and the length $L$ can be as short as
$100$\,nm~\cite{Chaste2011, Laird2012}. This expression is derived
for the fundamental mode of a rod (Eq.~S6 in the supplementary
information of Ref.~\cite{Eichler2012}), and $z_s$ is the
characteristic static displacement induced by the bending.
Assuming that $z_s$ ranges from $1$ to $10$\,nm, and using $L =
1.8$\,$\mu$m and the graphite density $\rho = 2300$\,kgm$^{-3}$
and Young modulus $E = 1$\,TP, we obtain $\beta =
3-30\cdot10^{24}$\,m$^{-1}$s$^{-2}$, which is consistent with the
value obtained from our measurements. The quadratic nonlinear
force associated to symmetry breaking is $3$ orders of magnitude
larger than the quadratic electrostatic force, $-\partial_z^3 C_g
(V_g^{dc})^2/2m\cdot \delta z^2(t)$. The observed decrease of
$\omega_0$ with the increasing resonant driving in Fig.~4c and d
indicates that the cubic nonlinear (Duffing) force has no
substantial effect on the dynamics of the resonator. This points
out that the actual static deformation $z_s$ is large compared to
the vibration amplitude (because the dynamical cubic restoring
force scales as $F_3\simeq F_2\cdot \delta z(t)/z_s$), thus
supporting our above assumption that $z_s=1-10$\,nm.

The observation of a peak in $I_{LF}$ for only about half of the
mechanical eigenmodes indicates that $\beta$ varies from one
eigenmode to the next. This is something expected from the
interplay between the shapes of the vibrational eigenmodes and the
static deformation along the nanotube if the static displacement
is primarily in one plane. Our data suggest that the static
displacement is essentially perpendicular to the gate electrode.
In such a geometry, the lowest-frequency eigenmode detected in
Fig.~4b corresponds to the lowest-energy mode vibrating
(essentially) parallel to the surface of the gate electrode, as
shown in Ref.~\cite{Eichler2012}. A static deformation of the
nanotube towards the gate electrode does not break the vibration
symmetry of this mode, because the elastic tension inside the
nanotube is equal for $+\delta z$ and $-\delta z$. As a result,
the amplitude of $I_{LF}$ should be weak, in agreement with the
measurements. The second eigenmode in Fig.~4b is assigned to the
lowest-energy mode vibrating in a direction (essentially)
perpendicular to the gate electrode~\cite{Eichler2012}. In the
presence of a static deformation towards the gate electrode, this
mode experiences symmetry breaking of vibrations. A peak shows up
in $I_{LF}$, as observed in Fig.~4a.

Having shown that symmetry breaking leads to motion at (nearly)
zero-frequency, we demonstrate other connections between symmetry
breaking and the mechanics of nanotube resonators. A hallmark of
nanotube resonators is that the resonance frequency can be widely
tuned with $V_g^{dc}$. Symmetry breaking is expected to control
this tunability in $\omega_0$ by an amount
\begin{equation}
\Delta\omega_0= \frac{\beta \partial_z
C_{g}}{2m\omega_0^3}(V_g^{dc})^2 \label{eq:variation-frequence1}
\end{equation}
($V_g^{dc}$ is here offset so that $V_g^{dc}=0$ when $\omega_0$ is
minimum). We estimate that
$m\simeq4$~ag assuming that the length of the nanotube is equal to
the trench width ($1.8$\,$\mu$m) and using the typical radius
($1.5$\,nm) obtained with our CVD recipe. Using the curvature of
$\Delta \omega_0 (V_g^{dc})$ near the minimum of $\omega_0$, we
get that $\beta = 3 (\pm 1) \cdot 10^{24}$\,m$^{-1}$s$^{-2}$,
which is close to the value estimated above. This result
underscores that symmetry breaking is connected to the response of
the resonance frequency to $V_g^{dc}$. We emphasize that
Eq.~(\ref{eq:variation-frequence1}) is only valid for not too large
$V_g^{dc}$, as it is the leading-order term of the expansion of
$\omega_0$ in $V_g^{dc}$. Here we find that Eq.~(\ref{eq:variation-frequence1}) applies in the
whole range of $V_g^{dc}$ that we studied, and that $\beta$ and
the bending of the nanotube both remain essentially constant
within this range. This suggests that the bending is a consequence
of the mechanical tension built in during the fabrication process.
In the future, it will be interesting to measure $\beta$ as a
function of $V_g^{dc}$ for other nanotube resonators.

In the presence of thermal vibrations, symmetry breaking leads to
spectral broadening~\cite{Dykman1971}. Because the amplitude of
thermal vibrations fluctuates in time, the nonlinearity-induced
shift in $\omega_0$ (Fig.~4d) also fluctuates and, therefore,
broadens the mechanical resonance. The broadening in $\omega_0$
reads
\begin{equation}
\label{eq:frequency_spread1} \overline{\delta\omega}= 5\beta^2
k_BT/6m\omega_0^5
\end{equation}
when the cubic restoring force is negligible compared to the
quadratic one (supplemental section X). Using $\beta= 4.3 \cdot
10^{24}$\,m$^{-1}$s$^{-2}$, we get
$\overline{\delta\omega}=2\pi\times7.5\cdot10^5$~Hz at room
temperature. This corresponds to an apparent quality factor of 67,
which is comparable to the value of $\simeq 50$ measured with the
FM technique. We emphasize that this broadening is analogous to
dephasing of two-level systems and qubits, which sets the
characteristic time $T_2$. The measured broadening is not related
to dissipation, so that the energy relaxation time could be much
longer than $1/\overline{\delta\omega}$ (in fact, it is in this
case that Eq.~(\ref{eq:frequency_spread1}) gives the spectral
broadening). For eigenmodes with a small $\beta$, the broadening
can be due to the cubic restoring force~\cite{Dykman1971}.
Mechanical resonances might be further broadened by the coupling
between eigenmodes~\cite{Dykman1971}, as shown by recent
simulations of nanotube resonators~\cite{Barnard2012}.

We assumed in our analysis of $I_{LF}$ that the response of the
amplitude of the vibrational motion is linear with the driving
force. When the response becomes nonlinear at large driving forces
due to the restoring force nonlinearity, the ratio of $I_{LF}$ at
$\omega_{AM}$ and $2\omega_{AM}$ is expected to deviate from 4.
Calculations show that the width of the peak in $I_{LF}$ remains
nearly constant upon varying the driving force, in contrast to the
measurements in Fig.~4c. A general theory that incorporates
nonlinearities in both the restoring force and damping
~\cite{Eichler2011,Dykman1975,Croy2012,Zaitsev2012,Lifshitz2008}
as well as thermal vibrations is beyond the scope of this Letter.
We note that our new technique to measure the motion of the
equilibrium position allows to study the response of the resonator
over a broad parameter range in driving force.

% %%%%%%%%%%%%%%%%%%%%%%%%%%%%%%%%%%%%%%%%%%%%%%%%%%%%%%%%%%%%%%%%%%%%%%%%
%\section{Summary}
% %%%%%%%%%%%%%%%%%%%%%%%%%%%%%%%%%%%%%%%%%%%%%%%%%%%%%%%%%%%%%%%%%%%%%%%%
In conclusion, we demonstrate that symmetry breaking leads to a
motion at nearly zero-frequency in response to resonant excitation
of the vibrations. Our results indicate that symmetry breaking of
vibrational modes also leads to such important dynamical
properties as the apparent low quality factor of nanotube
resonators at 300~K, and the shift of the vibration frequency in
response to both (i) the static gate voltage and (ii) the
amplitude of the oscillating driving force. A future strategy to
improve the apparent $Q$ at 300~K is to tune $\beta$ with the gate
voltage in order to compensate the spectral broadening due to
symmetry breaking with that due to the Duffing nonlinearity.
Symmetry breaking is important for other vibrational systems of
current interest, such as graphene
resonators~\cite{Eichler2011,Bunch2007,Chen2009,Singh2010,Song2012,Reserbat2012}
and levitating particles~\cite{Chang2009,Raizen2011,Quidant2012}.
Our new technique may help to reveal this effect in such systems.
Symmetry breaking also leads to mode mixing and to parametric
resonance in response to additive driving. This holds promise for
a number of applications, such as controlled mode
mixing~\cite{Yamaguchi2012,Antonio2012,Faust2012} and phase noise
cancelation~\cite{Dykman1990,Kenig2012,Villanueva2011}.

\vspace{20pt} \textbf{Acknowlegements}\\ We acknowledge support
from the European Union through the RODIN-FP7 project, the
ERC-carbonNEMS project, and a Marie Curie grant (271938), the
Spanish state (FIS2009-11284), the Catalan government (AGAUR,
SGR), and the US Army Research Office.\vspace{20pt}\\
\textbf{Author contributions}\\ A.E. fabricated the devices and
carried out the measurements. J.M. participated in the
measurements. M.I.D. provided support with the theory and wrote
the theoretical part of Supplementary Information. All the authors
contributed to writing the manuscript. M.I.D. and A.B. conceived
the experiment. A.B. supervised the work.\vspace{20pt}\\
Correspondence and requests for materials should be addressed to
A.B. (adrian.bachtold$@$icfo.es)

\newpage
\begin{figure}
\includegraphics[width=125mm]{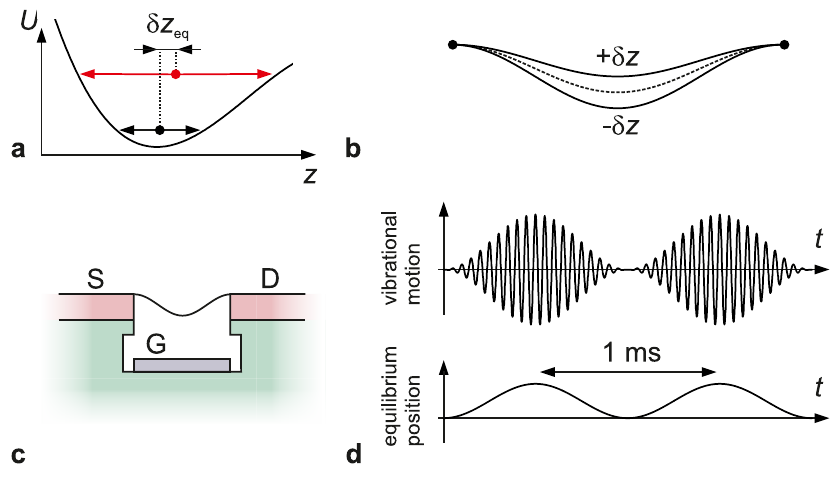}% size for 2column
\caption{\label{Figure 1} \textbf{\textbar~Effect of curvature in
a nanotube resonator.} \textbf{a}, Symmetry breaking of the
restoring potential $U(z)$. The equilibrium position depends on
the energy of the resonator mode. \textbf{b}, Schematic of a
curved resonator. The dashed line represents the static profile of
the resonator, that is, when it does not vibrate. The plain lines
show the profiles for $+\delta z$ and $-\delta z$. \textbf{c}, The
resonator studied consists of a carbon nanotube suspended over a
trench between source (S) and drain (D) electrodes. A gate
electrode (G) is defined at the bottom of the trench. The trench
has a width of $1.8$\,$\mu$m and a depth of $\sim 350$\,nm.
\textbf{d}, The vibrational motion is amplitude modulated at
$\omega _{AM}$ (upper schematic). As a result, the equilibrium
position is modulated with the same period (lower schematic).}
\end{figure}

\begin{figure}
\includegraphics[width=125mm]{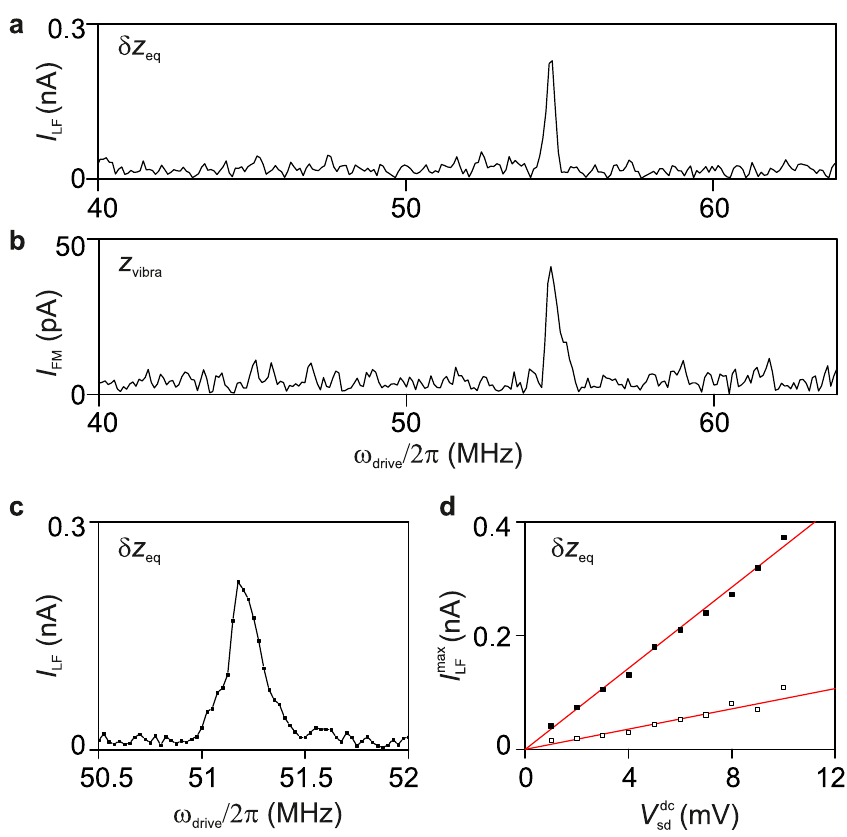}% size for 2column
\caption{\label{Figure 2} \textbf{\textbar~Characterization of the
low-frequency current.} \textbf{a}, $I_{LF}$ as a function of
$\omega_{drive}$. $V_g^{ac} = 0.53$\,mV, $V_{sd}^{dc} = 10$\,mV,
and $V_g^{dc} = -0.45$\,V. \textbf{b}, Mechanical vibrations
detected with the FM technique~\cite{Gouttenoire2010}. $V_{sd}^{ac} =
1.1$\,mV and $V_g^{dc} = -0.45$\,V. \textbf{c}, $I_{LF}$ versus
$\omega_{drive}$ with $V_g^{ac} = 1.1$\,mV, $V_{sd}^{dc} =
10$\,mV, and $V_g^{dc} = -0.4$\,V. \textbf{d}, $I_{LF}^{max}$ as a
function of $V_{sd}^{dc}$ measured at $\omega_{AM}$ (black
squares) and $2\omega_{AM}$ (open squares). $V_g^{ac} = 2.2$\,mV
and $V_g^{dc} = -0.4$\,V.}
\end{figure}

\begin{figure}
\includegraphics[width=125mm]{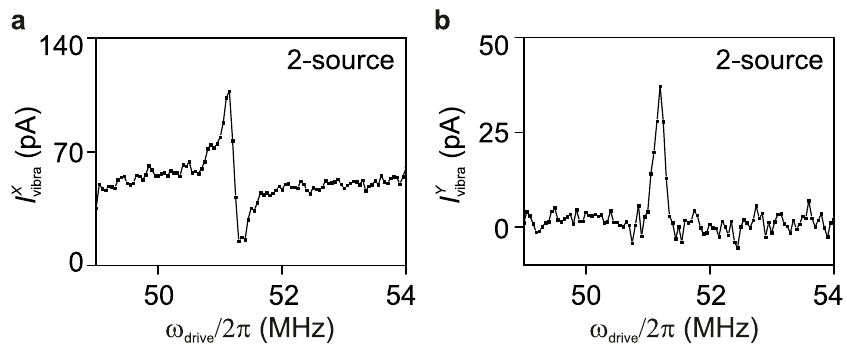}% size for 2column
\caption{\label{Figure 3} \textbf{\textbar~Vibrational motion
measured with the 2-source mixing technique.} \textbf{a,} $X$
quadrature and \textbf{b,} $Y$ quadrature of the current measured
with the lock-in amplifier. $I_{vibra}^{X}$ consists of a current
proportional to the real part of the vibrational amplitude in
addition to a purely electrical background current.
$I_{vibra}^{Y}$ is proportional to the imaginary part of the
vibrational displacement. $V_g^{ac} = 1.1$\,mV, $V_{sd}^{ac} =
0.3$\,mV, and $V_g^{dc} = -0.4$\,V.}
\end{figure}

\begin{figure}
\includegraphics[width=125mm]{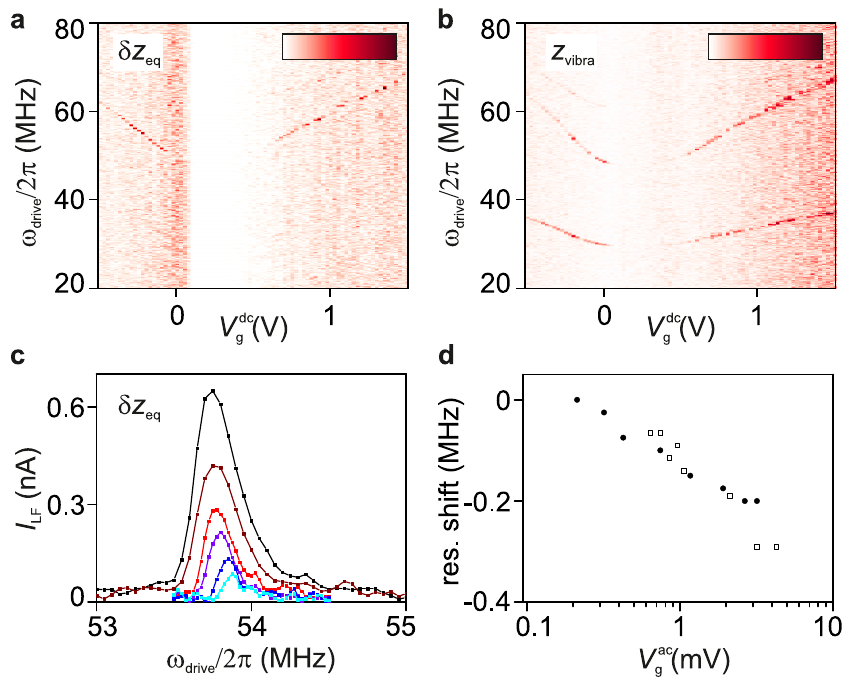}% size for 2column
\caption{\label{Figure 4} \textbf{\textbar~Response of the low
frequency current to static and oscillating forces.} \textbf{a},
$I_{LF}$ as a function of $\omega_{drive}$ and $V_g^{dc}$.
$V_g^{ac} = 0.53$\,mV, $V_{sd}^{dc} = 10$\,mV. Color bar: $0$
(white) to $280$\,pA (dark red). The background signal varies with
$V_g^{dc}$; this variation likely has a purely electrical origin.
The number of measurement points is kept as low as possible so
that resonances are captured with about 3 points along the
frequency axis. \textbf{b}, Current as a function of
$\omega_{drive}$ and $V_g^{dc}$ measured with the FM technique.
$V_{sd}^{ac} = 2.2$\,mV. Color bar: $0$ (white) to $170$\,pA (dark
red). \textbf{c}, Measured lineshapes of $I_{LF}$ as a function of
$\omega_{drive}$ for different $V_g^{ac}$. $V_g^{ac} = 4.2$,
$3.5$, $2.5$, $1.6$, $1$, and $0.6$\,mV, from top to bottom.
\textbf{d}, Resonance shift extracted from the data in \textbf{c}
(black dots) and from FM measurements (open squares).}
\end{figure}

\clearpage

\large
\begin{center}

\textbf{Supplementary Material for: Symmetry breaking in a
mechanical oscillator made from a carbon nanotube}

\normalsize

\vspace{5 mm}

A. Eichler$^{1,2}$, J. Moser$^{1,2}$, M. I. Dykman$^{3}$, and A. Bachtold$^{1,2}$

\textit{$^1$ICFO - Institut de Ciencies Fotoniques, Mediterranean
Technology Park, 08860 Castelldefels, Barcelona, Spain,}

\textit{$^2$Institut Catal{\`a} de Nanotecnologia, Campus de la
UAB, E-08193 Bellaterra, Spain, and}

\textit{$^3$ Department of Physics and Astronomy, Michigan State
University, East Lansing, Michigan 48824, USA}

\end{center}

\small

\section{General expression for the low-frequency conductance}

We consider the conductance of a suspended
nanotube (supplementary references~\cite{Zant_book,Moser_book}) in the presence of a gate
voltage that has a large DC component and a small AC component at
a frequency close to the frequency of the eigenvibrations of the
nanotube.

%\subsection{AC conductance of the vibrating nanotube}

We assume that $(i)$ the nanotube conductance is a function only
of the total charge $q$ of the nanotube, $(ii)$ the charge
distribution along the nanotube is independent of the gate voltage
$V_g$, and $(iii)$ the system is in the adiabatic limit, i.e. the
vibration dynamics is much slower than the electron dynamics. Then
$q$ is related to $V_g$ by the capacitance $C_g$. We consider the
effect on conductance of the bending mode of the nanotube which is
polarized in the direction $z$ perpendicular to the gate.

Based on the above assumptions we write the conductance as
\begin{equation}
\label{eq:conductance_general_sym} G(q(t))\simeq
G(q_{0})+\partial_q G\,\delta q(t) + \frac{1}{2}\partial^2_qG[\delta q(t)]^2+ \ldots\,.
\end{equation}
For the temperatures used in our experiments, where the Coulomb
blockade effect is insignificant, the term $\propto \delta q^2$ is
comparatively small. In what follows we will disregard it.
Incorporating this term will not change the qualitative results
(see section VII).

The charge increment $\delta q(t)$ is a function of the
time-dependent (AC) increment of the gate voltage $\delta
V_{g}(t)$ and the (AC) vibrational displacement $\delta z(t)$,
which is the displacement of the nanotube at the antinode of the
vibrational mode with the largest amplitude, for a given mode. For
small $|\delta V_g|$ and $|\delta z|$
\begin{equation}
\label{eq:deltaQ_general_sym} \delta q(t)\simeq \partial
_{V_g}q\,\delta V_{g}(t)+\partial_z
q\,\delta z(t)+\frac{1}{2}\partial^2_{V_g}
q\,\delta V_{g}(t)^2+
\partial_{z}\partial_{V_g}
q\,\delta z(t)\,\delta V_g(t)+\frac{1}{2}\partial^2_z q\,\delta z(t)^2\,.
\end{equation}
The coefficients in this expression have a simple form in the case where the charge $q$ is related to the gate voltage by the gate capacitance $C_g$, which itself depends on the displacement of the nanotube. We then have
\begin{equation}
\label{eq:coefficients}
\partial_{V_g}q=C_g,\qquad \partial_z q = V_g^{dc}\partial_z C_g,\qquad \partial_{z}\partial_{V_g}q = \partial_zC_g,\qquad \partial^2_z q=V_g^{dc}\partial^2_z C_g
\end{equation}

whereas we can set $\partial^2_{V_g}q =0$, assuming the capacitor to be linear. In Eq.~(\ref{eq:coefficients}) $V_g^{dc}$ is the DC gate voltage, which is assumed to be large compared to $\delta V_g(t)$.

We will consider an AC modulation $\delta V_g(t)$ (with amplitude $V_g^{ac}$) at frequency $\omega_{drive}$ close to the eigenfrequency of the nanotube $\omega_0$. If this modulation is not too weak, the major contribution to the AC displacement $\delta z$ is the one induced by this resonant modulation whereas the thermal displacement can be disregarded.

The quantity of immediate interest to us is the quasi-static change of the conductance in response to $\delta V_g$. As explained in the main text, to detect this change we consider a periodic signal with  slowly modulated amplitude,
\begin{equation}
\label{eq:modulation}
\delta V_g(t) = V_g^{ac}(t)\cos\omega_{drive} t, \qquad V_g^{ac}(t)=V_0(1-\cos(\omega_{AM}t)), \qquad\omega_{AM}\ll \omega_{drive}.
\end{equation}
There are several contributions to the low-frequency response of the conductance to the modulation (\ref{eq:modulation}). To study them we first estimate the response of the resonator to the modulation assuming that the resonator dynamics is linear, $\delta z = \delta z^{lin}$. The linearized equation of motion in the simplest case of viscous friction reads
\begin{equation}
\delta \ddot z^{lin}+2\Gamma \delta \dot z^{lin}+\omega_{0}^{2}\delta z^{lin}=\frac{F_d(t)\cos(\omega_{drive}t)}{m}\,, \qquad F_d(t)=\partial_{z}C_{g} V_{g}^{dc} V_{g}^{ac}(t) \label{eq:harmonic_osc}
\end{equation}
where $m$ is the mass of the nanotube, $2\Gamma=\omega_{0}/Q$ is the decay rate of the oscillator with
quality factor $Q$, and $F_d(t)$ is the AC driving force amplitude. In what follows we assume that the frequency of the amplitude modulation, $\omega_{AM}$, is small compared to the decay rate $\Gamma$, so that the induced vibrations adiabatically follow $V_g^{ac}(t)$. Then
\begin{align}
\label{eq:Lorentz_amplitude_sym}
\delta z^{lin}(t)=A_{}^{lin}(\omega_{drive},t)&\cos(\omega_{drive}t-\phi), \qquad A^{lin}(\omega_{drive},t)=\frac{F_d(t)/m}{\sqrt{\left(\omega_{0}^{2}-\omega_{drive}^{2}\right)^{2}+4\Gamma^{2}\omega_{drive}^{2}}}, \nonumber \\ &\phi=\arctan\left(\frac{2\Gamma\omega_{drive}}{\omega_{0}^{2}-\omega_{drive}^{2}}\right)\,.
\end{align}

From Eqs.(\ref{eq:modulation}), and (\ref{eq:Lorentz_amplitude_sym}) it follows that all terms that have a quadratic dependence on $\delta V_g$ and $\delta z$ in Eq.~(\ref{eq:deltaQ_general_sym}) have a slowly varying part, which oscillates with period $2\pi/\omega_{AM}$. If the distance between the gate electrode and the nanotube is $h$ (it is of the order of the depth of the trench, $\sim 350$\,nm), then $\partial_z C_g\sim C_g/h$. Then in the linear approximation the amplitude on resonance is $A_{res} = F_d / 2 m \Gamma \omega_0 \sim C_g V_g^{ac} V_g^{dc}/m \Gamma \omega_0 h$. The slowly varying parts of the last two terms in Eq.~(\ref{eq:deltaQ_general_sym}) are
\begin{equation}
\label{eq:two_terms}
\sim C_g V_g^{ac} A^{lin}/h, \qquad  C_g V_g^{dc}(A^{lin})^2/h^2.
\end{equation}
A simple estimate shows that, for the device parameters, the
second term is much larger than the first for resonant driving,
which means that the term $\propto \delta V_g\delta z$ in
Eq.~(\ref{eq:deltaQ_general_sym}) should be disregarded in the
analysis of the low-frequency conductance.

\section{Nonlinear response of the vibrational mode}

In the linear approximation
[Eq.~(\ref{eq:Lorentz_amplitude_sym})], the term $\propto \delta
z(t)$ in the expression for the charge and thus the conduction
modulation [Eq.~(\ref{eq:deltaQ_general_sym})] are oscillating at
high frequencies $\omega_{drive}$, $\omega_{drive}\pm\omega_{AM}$.
However, the vibrations of the nanotube are nonlinear, and this
leads to the onset of slowly varying terms in the displacement
$\delta z(t)$. To find these terms we write the part of the
capacitive energy and the internal energy of the mode that is
nonlinear in $\delta V_g$ and $\delta z$,
\begin{equation}
\label{eq:nonlinear_energy}
H_c^{nl}=-\frac{1}{2}\partial_zC_g\delta V_g^2\,\delta z
-\frac{1}{2}\partial^2_zC_g V_g^{dc}\delta V_g\delta z^2 +
\frac{1}{3}m\beta\delta z^3 + \frac{1}{4}m\gamma\delta z^4 +
\ldots.
\end{equation}
The first two terms in this expression describe the nonlinear capacitive energy, whereas the last two terms refer to the nonlinear part of the vibrational energy. We emphasize that the term which is cubic in $\delta z$ is present only because the mode lacks inversion symmetry: this term is the indication of symmetry breaking (it corresponds to a force that is quadratic in $\delta z$). Such symmetry breaking may result from the gate voltage which bends the nanotube. Therefore we expect that $\beta$ depends on $V_g^{dc}$. On the other hand, the term $\propto \gamma$ is the familiar Duffing nonlinearity, which has been known to play an important role in the vibrational dynamics of nanotubes (supplementary references~\cite{Zant_book,Moser_book}).%Eichler2011, Eichler2012}.

We emphasize again that $\delta z$ refers to the maximal
displacement for the considered mode in the $z$-direction, i.e.,
toward the gate electrode. More generally, for bending modes, one
should think of the displacement $\delta{\bf r}$ as a function of
length $l$ along the nanotube ($\delta{\bf  r}$ locally transverse
to $d{\bf l}$). Then, for example, the term that leads to
$(m/3)\beta \delta z^3$ in Eq.~(\ref{eq:nonlinear_energy}) would
be written as a triple integral over the length
\[H_{sym-brk}=\frac{1}{3}m\tilde\beta\int dl_1dl_2dl_3 f_{ijk}(l_1,l_2,l_3)\delta r_i\delta r_j\delta r_k.\]
Function $f$ here is nonzero only for a nanotube with broken
symmetry, i.e., where the energy changes if one replaces
$\delta{\bf r}\to -\delta{\bf r}$. The term $\propto\delta z^3$ in
Eq.~(\ref{eq:nonlinear_energy}) is obtained if one substitutes
$\delta {\bf r}(l)$ with the solution of the harmonic problem, for
the considered mode. In general, in nanotubes with broken
symmetry, the coupling between different modes leads to an energy
that is cubic in the displacements of the modes.

A simple calculation shows that, to leading order, the first 3 terms in Eq.~(\ref{eq:nonlinear_energy}) give the slowly varying terms in $\delta z(t)$ of the form
\begin{equation}
\label{eq:slow_displacement} \delta z_{slow}(t)\approx \partial_z
C_g\frac{V_g^{ac}(t)^2}{4m\omega_0^2} +\partial^2_z C_g\,V_g^{dc}
\frac{V_g^{ac}(t)A^{lin}(\omega_{drive},t)}{2m\omega_0^2}\cos\phi
- \beta\frac{A^{lin}(\omega_{drive},t)^2}{2\omega_0^2}
\end{equation}
Here, the first term is very much smaller than
the second term for typical device parameters; the ratio of these
terms is of the same order of magnitude as the ratio of the terms
in Eq.~(\ref{eq:two_terms}). We note, however, that on exact
resonance, $\omega_{drive}=\omega_0$, and we have $\cos\phi =0$.
Therefore either $|\delta z_{slow}|$ displays an extremely narrow
and extremely deep dip as a function of $\omega_{drive}$, which is
expected for $\beta \to 0$, or the dominating term in
Eq.~(\ref{eq:slow_displacement}) is the last term, which comes
from the broken inversion symmetry.

It is necessary also to look at the ratio of the contributions to
the conductance modulation of the second term in $\delta z_{slow}$
in Eq.~(\ref{eq:slow_displacement}) and the term $\partial^2_z q
\delta z^2$ in Eq.~(\ref{eq:deltaQ_general_sym}). One can easily
see that this ratio is $\sim \Gamma/\omega_0=(2Q)^{-1} \ll 1$.
Therefore the leading-order contribution to the scaled slowly
varying conductance is
\begin{equation}
\label{eq:major_contributions} \delta G \approx
\partial_q G V_g^{dc} \overline{\delta z(t)^2}\left(
\frac{1}{2}\partial^2_zC_g  -
\partial_zC_g\frac{\beta}{\omega_0^2}\right).
\end{equation}
Here, bar means averaging over the period of fast oscillations $2\pi/\omega_{drive}$.

Experimentally, the easiest way to separate the two contributions
to $\delta G$ in Eq.~(\ref{eq:major_contributions}) is by
estimating $\partial_z C_g$, $\partial^2_z C_g$, and $\beta$ from
independent measurements. In the following sections, we
demonstrate how this estimate is done for our device. The estimate
indicates that the first term in the bracket in
Eq.~(\ref{eq:major_contributions}) is too small to account for our
measurements. We also estimate $\beta$ by analyzing the shift of
$\omega_0$ as a funtion of $z_{vibra}$ (the amplitude of $\delta
z(t)$). We find that this latter estimate is in good agreement
with our measurement. Therefore, the major effect is coming from
the symmetry breaking of the vibrations. In the main text and in
the following, we refer to the slow motion $\delta z_{slow}$ in
terms of a (quasi-static) shift of the equilibrium position,
$\delta z_{eq}$.

\section{Electrical conductance and capacitance of the nanotube}
In Figure~5a, we show the electrical conductance $G$ of the
nanotube device presented in the main text as a function of the
constant gate voltage $V_g^{dc}$ at a temperature of $65$\,K. We
find this trace to be reproducible over a timescale of weeks (a
current annealing procedure is performed every day to counter the
effects of contamination with residual gas particles). The
conductance of a nanotube depends on its charge carrier density,
which is controlled by $V_g^{dc}$. The voltage couples to the
nanotube through the capacitance $C_g$, which we can easily
determine: in the Coulomb blockade regime, the separation between
two conductance peaks is given by $\Delta V_g = e C_g$, where $e$
is the electron charge. From the measurement in Fig.~5b, we get
$C_g = 12$\,aF, which is in agreement with an estimation based on
the device geometry:
\begin{equation}
\label{eq:capacitance}
C_g = \frac{2 \pi \epsilon_0 L}{\ln(2 d / r)}.
\end{equation}
Here, $\epsilon_0$ is the vacuum permittivity, $L = 1.8$\,$\mu$m
is the nanotube length, and $d = 350$\,nm is the equilibrium
distance between the nanotube and the gate electrode. Since we
cannot measure the diameter of the nanotube due to the large
surface roughness of the electrodes in the studied device, we use
a typical value for the radius ($r = 1.5$\,nm). We expect that the
capacitance weakly depends on temperature, as it is determined
primarily by geometrical factors. We determine $\partial_z C_g$
and $\partial^2_z C_g$ by differentiating
Eq.~(\ref{eq:capacitance}) and get $\partial_z C_g = 5.6$\,pF/m
and $\partial^2_z C_g = 21$\,$\mu$F/m$^2$.

From the measurements of the resonance frequency as a function of
$V_g^{dc}$ in Fig.~4b of the main text, we obtain a voltage offset
of $0.45$\,V, which corresponds to the work function difference
between the nanotube and the gate electrode. This offset in
$V_g^{dc}$ is included in all the estimates. However, the values
of $V_g^{dc}$ that we indicate in the main text and the
supplementary information are always the voltages that are applied
to the gate electrode.

\begin{figure}
\includegraphics[width=125mm]{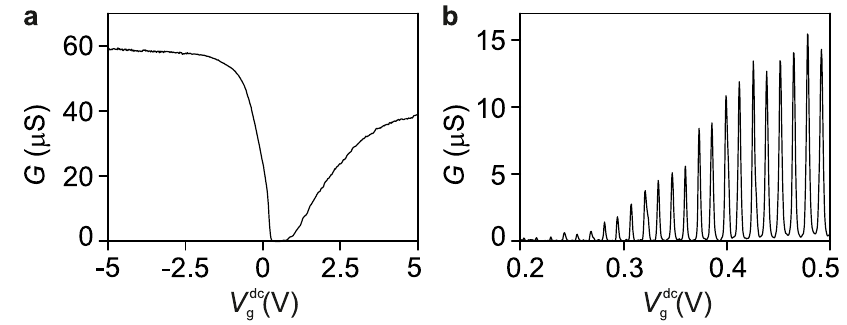}% size for 2column
\caption{\label{Figure S1} \textbf{Nanotube conductance as a
function of static gate voltage.} \textbf{a,} Nanotube conductance
at $65$\,K as a function of the constant gate voltage $V_g^{dc}$
applied to the gate electrode. \textbf{b,} Nanotube conductance at
$650$\,mK. The device is in the Coulomb blockade regime. The
spacing between consecutive conductance peaks is $\Delta V_g= e
C_g$, where $e$ is the electron charge.}
\end{figure}

\section{Measurements of vibrational motion}
We discuss first the frequency mixing (FM)
technique~\cite{Gouttenoire2010}. A driving voltage $V_{sd}^{ac}$
is applied to the source electrode. Modulating the frequency (with
a modulation rate of $671$\,Hz and a frequency deviation of
$100$\,kHz) results in a current ($I_{FM}$) at $671$\,Hz. The gate
electrode is biased with $V_g^{dc}$ to tune the resonance
frequencies. This technique has a low current background and is
typically more sensitive than the 2-source technique, so we
preferentially use it to detect the eigenmodes of a nanotube
resonator. The main drawback of the FM technique is that the
measured signal is not proportional to $z_{vibra}$, but to the
derivative with respect to the frequency of the real part of the
displacement.

In the 2-source technique~\cite{Sazonova2004}, we apply a driving
voltage $V_g^{ac}$ to the gate in addition to a DC voltage
$V_g^{dc}$. The motion of the nanotube is detected by applying a
second, smaller voltage $V_{sd}^{ac}$ to the source. The two
oscillating voltages are slightly detuned, and the mixing current
$I_{mix}$ is measured at the detuning frequency (typically $\delta
\omega / 2 \pi = 10$\,kHz). When the displacement is written as
$z(t)={\rm Re}\left[\tilde{z}(\omega)\right]\cos(\omega t)+{\rm
Im}\left[\tilde{z}(\omega)\right]\sin(\omega t)$, the mixing
current $I_{mix}$ measured with the 2-source technique has the
form~\cite{Eichler2011B}
\begin{align}
I_{mix} =& \frac{1}{2} V_{sd}^{ac} \partial_{V_g} G\left(V_g^{ac} \cos(\delta \omega t - \varphi_E)+ V_g^{dc} \frac{\partial_z C_g}{C_g} {\mathrm Re}\left[\tilde{z}(\omega)\right] \cos(\delta \omega t - \varphi_E) \right. \nonumber \\
& \left. + V_g^{dc} \frac{\partial_z C_g}{C_g}
{\mathrm Im}\left[\tilde{z}(\omega)\right] \sin(\delta \omega
t - \varphi_E) \right) \label{mixingcurrent}
\end{align}
where $G$ is the conductance of the nanotube, and $\varphi_E$ is
the phase difference between the voltages applied to source and
gate. For a properly tuned phase of the lock-in amplifier, the
out-of-phase component of the lock-in amplifier output, $Y$,
corresponds to the imaginary part of the resonant displacement
[third term in Eq.~(\ref{mixingcurrent})], whereas the in-phase
component, $X$, corresponds to the real part of the resonant
displacement [second term in Eq.~(\ref{mixingcurrent})] added to a
background (first term in Eq.~\ref{mixingcurrent}) that weakly
depends on frequency near resonance with a given mode (we note
that it can have contributions from other modes). For the
modulation frequency $\omega$ close to resonance, we get for the
$Y$-component of the mixing current $I_{mix}$, which we denote as
$I_{vibra}^Y$,
\begin{align}
I_{vibra}^{Y} = \frac{1}{2} \cdot V_{sd}^{ac} V_g^{dc} \cdot \partial_{V_g} G \cdot \frac{\partial_z C_{g}}{C_g} \cdot z_{vibra},  \label{eq:mixing_resonant}
\end{align}
where $z_{vibra}$ is the amplitude of resonant forced vibrations.
For the considered small $V_g^{ac}$, $z_{vibra}$ is proportional
to the amplitude of $V_g^{ac}$. $I_{vibra}^{Y}$ can be
conveniently read out from the measurement.

\section{Estimation of vibration amplitude}

Equation~(\ref{mixingcurrent}) allows estimating the vibration
amplitude of the resonator for resonant driving by comparing the
out-of phase current on resonance, $I_{vibra}^{Y}$, to the
background far from resonance,
$I_{off-res}^{X}$~\cite{Sazonova2004, Eichler2011B}. Using
Eq.~(\ref{eq:capacitance}), we get that
\begin{equation}
z_{vibra} \simeq d \cdot \ln\left(\frac{2 d}{r} \right) \frac{I_{vibra}^{Y}}{I_{off-res}^{X}} \frac{V_g^{ac}}{V_g^{dc}}
\end{equation}

The measurement in Fig.~3 yields a value of $z_{vibra} = 2.1$\,nm
for the modulation amplitude $V_g^{ac}=1.1$~mV.

\section{Detection of the motion of the
equilibrium position} In the following, we explain in more detail
the technique we develop to detect the motion of the equilibrium
position of the nanotube resonator due to symmetry breaking. We
drive the resonator with an amplitude modulated (AM) driving
force, which causes an AM vibrational motion (Eq.~2 of the main
text). The amplitude change of the vibration is quasi-adiabatic
from the point of view of the resonator (we checked that the
result is independent of the modulation period $2\pi/\omega_{AM}$
up to 0.1 ms.) In the presence of AM modulation the quadratic
nonlinear force $F_2 \propto \delta z(t)^2$ leads to the
oscillation of the nanotube equilibrium position, $\delta z_{eq}$,
as illustrated in Fig.~1d of the main text.

The conductance $G$ of the nanotube depends both on the voltage
that is applied to the gate electrode ($V_g$) and on the
capacitance between the gate electrode and the nanotube ($C_g$).
We can rewrite Eqs.~6-8 for the change in conductance as
\begin{align}
\delta G(t) = \partial_{V_g} G \cdot \delta V_g(t) + \partial_{C_g} G \cdot \delta C_g(t).
\end{align}
In the analysis of the low-frequency conductance the first term on
the right hand side can be neglected because the voltage that we
apply to the gate electrode has no term at the frequency of
interest $\omega_{AM}/ 2 \pi$ (we have verified this using a
signal analyzer). Taking into account the analysis of Sec.~I, we
then write
\begin{align}
\delta G(t) = \partial_{V_g} G \cdot V_g^{dc} \frac{\partial_z C_g}{C_g} \delta z(t).
\end{align}
The slow oscillation of the conductance is caused by the motion of
the equilibrium position due to symmetry breaking. As shown in
Fig.~1d of the main text, for comparatively weak resonant
modulation, where the AM vibrational motion is of the form of
$\delta z(t) = z_{vibra} \cdot \cos(\omega_{drive} t) \cdot [1 -
\cos(\omega_{AM} t)]$, the equilibrium position $\delta
z_{eq}\propto [\overline{\delta z(t)^2}]$ oscillates with period
$2\pi/\omega_{AM}$. The vibrations are nonsinusoidal,
\begin{align}
\delta z_{eq}(t) =\delta z_{eq}^{0}\left[-\cos\omega_{AM}t
+\frac{3}{4}+\frac{1}{4}\cos 2\omega_{AM}t\right].
\end{align}
The amplitude of the low-frequency current oscillation measured at
$\omega_{AM}$ is then simply
\begin{align}
I_{LF}= V_{sd}^{dc} \cdot \delta G = V_{sd}^{dc}
V_g^{dc}\left|\partial_{V_g} G \cdot \frac{\partial_z C_g}{C_g}
\right|\delta z_{eq}^{0}. \label{AM_current}
\end{align}

This current corresponds to the second term in the bracket of
Eq.~(\ref{eq:major_contributions}). By comparing
Eq.~(\ref{AM_current}) to Eq.~(\ref{eq:mixing_resonant}), we can
express the low-frequency current at resonance as
\begin{align}
I_{LF}^{max} =2 I_{vibra}^{Y} \cdot \frac{\delta
z_{eq}^{0}}{z_{vibra}} \cdot \frac{V_{sd}^{dc}}{V_{sd}^{ac}}
\label{I_vibra_to_I_LF}
\end{align}
provided that $I_{LF}$ and $I_{vibra}^{Y}$ are measured with the
same driving force. (The driving force comes from the gate voltage
oscillation at frequency $\omega$; however, in the 2-source
technique used to measure $I_{vibra}^Y$ the voltage amplitude is
not modulated; we note again that we study the regime where
$V_g^{ac}$ is comparatively small.)

\section{Estimation of the low-frequency currents due to the
capacitive nonlinearity and the conductance nonlinearity}

In this section, we first estimate the current that is expected
due to the nonlinearity of the capacitance with respect to
displacement, $\partial^2_z C_g$ [see
Eqs.~(\ref{eq:deltaQ_general_sym}), (\ref{eq:coefficients}), and
(\ref{eq:major_contributions})]. From
Eqs.~(\ref{eq:deltaQ_general_sym}) and (\ref{eq:coefficients}),
the current at $\omega_{AM}$ that we expect due to the capacitive
nonlinearity in the presence of a DC bias voltage ($V_{sd}^{dc}$)
has the form
\begin{align}
\label{eq:capacitive_nonlinearity1} I_{LF}^{capa} = V_{sd}^{dc}
\cdot \delta G = \frac{1}{2} V_{sd}^{dc} V_g^{dc} \cdot
\partial_{V_g} G \cdot \frac{\partial_z^2 C_g}{C_g} z_{vibra}^2.
\end{align}
Comparing Eq.~(\ref{eq:capacitive_nonlinearity1}) to
Eq.~(\ref{eq:mixing_resonant}), we can express this current in
terms of the 2-source mixing current measured on resonance in
Fig.~3 of the main text. We get
\begin{align}
\label{eq:capacitive_nonlinearity2} I_{LF}^{capa} = I_{vibra}^Y
\cdot \frac{V_{sd}^{dc}}{V_{sd}^{ac}} \cdot \frac{\partial_z^2
C_g}{\partial_z C_g} z_{vibra}.
\end{align}
where the currents are taken at the resonance frequency and for
the same amplitude $V_{g}^{ac}$. This relation is useful, because
it depends on a small number of parameters. Using $I_{vibra}^{Y} =
37$\,pA from Fig.~3b, $V_{sd}^{dc} = 10$\,mV, $V_{sd}^{ac} =
0.3$\,mV, $\partial_z C_g=5.6$\,pF/m, $\partial_z^2 C_g =
21$\,$\mu$F/m$^2$, and $z_{vibra} = 2.1$\,nm, we get that
$I_{LF}^{capa} = 9.7$\,pA$\pm 5$\,pA, which is far below $I_{LF} =
219$\,pA that we measured in Fig.~2c.

Next, we estimate the current that is expected due to the
nonlinearity of the conductance with respect to the gate voltage,
$\partial_{V_g}^2 G$, which is described by the last term in
Eq.~(\ref{eq:conductance_general_sym}). From
Eqs.~(\ref{eq:deltaQ_general_sym}) and (\ref{eq:coefficients}),
the current at $\omega_{AM}$ is
\begin{align}
\label{eq:conductance_nonlinearity1} I_{LF}^{cond} = \frac{1}{2}
V_{sd}^{dc} (V_g^{dc})^2 \cdot
\partial_{V_g}^2 G \cdot (\partial_z C_g/C_g)^2 z_{vibra}^2.
\end{align}
Using $V_{sd}^{dc} = 10$\,mV, $V_{g}^{dc} = -0.4$\,V, $\partial_z
C_g=5.6$\,pF/m, $C_g = 12$\,aF, and $z_{vibra} = 2.1$\,nm,
$\partial_{V_g}^2 G= 54 \mu$S/V$^2$, we get that $I_{LF}^{cond} =
0.19$\,pA.

\section{Estimation of the symmetry breaking strength $\beta$}

In the presence of the AM vibrational motion of the form of
$\delta z(t) = z_{vibra} \cdot \cos(\omega_{drive} t) \cdot [1 -
\cos(\omega_{AM} t)]$ and in the limit of small $\delta
z_{eq}^{0}$ (such that the restoring force can be approximated by
the spring force $m \omega_0^2$), $\delta z_{eq}^{0}$ is related
to the parameter $\beta$ as
\begin{align}
\omega_0^2 \delta z_{eq}^{0} = |\beta|
z_{vibra}^2\,. \label{eq:equilibrium_shift}
\end{align}
Using Eq.~(\ref{eq:equilibrium_shift}) together with
Eq.~(\ref{I_vibra_to_I_LF}), we arrive at the relation
\begin{align}
I_{LF}^{max} = 2 I_{vibra}^{Y} \cdot
\frac{V_{sd}^{dc}}{V_{sd}^{ac}} \cdot \frac{|\beta|}{\omega_0^2}
\cdot z_{vibra}. \label{eq:equilibrium_current}
\end{align}
Here, again the currents are taken at the resonance frequency and
for the same amplitude $V_{g}^{ac}$. This relation is useful,
because it depends on a small number of parameters that, in
addition, are well characterized. With $I_{vibra}^{Y} = 37$\,pA
from Fig.~3b, $V_{sd}^{dc} = 10$\,mV, $V_{sd}^{ac} = 0.3$\,mV,
$\omega_0 = 2 \pi \cdot 51$\,MHz, $z_{vibra} = 2.1$\,nm, and
$I_{LF}^{max}=I_{LF}^{0} = 219$\,pA measured in Fig.~2c of the
main text, we get $\delta z_{eq}^{0}=0.18$\,nm and $\beta =
4.3~\cdot~10^{24}$\,m$^{-1}$s$^{-2}$.

An alternative estimation of $\beta$ is possible using the fact
that the symmetry breaking force leads to a shift of the resonance
frequency with driving force. We have measured the shift of the
resonance frequency as a function of the vibrational amplitude
with the FM technique as well as with our new detection technique.
The values measured with the two methods agree well and are
plotted in Fig.~4d of the main text. We use the relation
(supplementary reference \cite{Landau_mech})
\begin{align}
\Delta \omega_0 = \frac{3}{8} \frac{\gamma_{eff}}{\omega_0} z_{vibra}^2,
\label{eq:backbone}
\end{align}
where $\Delta \omega_0$ is the shift of the resonance frequency and
\begin{align}
\gamma_{eff} = \gamma - \frac{10}{9} \omega_0^{-2} \beta^2 \label{eq:alpha_eff}
\end{align}
[$\gamma$ is the coefficient of the Duffing nonlinearity, see
Eq.~(\ref{eq:nonlinear_energy}).] We estimate $\gamma_{eff} = -
1.8 \cdot 10^{32}$\,m$^{-2}$s$^{-2}$ from Eq.~(\ref{eq:backbone})
by inserting $z_{vibra} = 2.1$\,nm, $\omega_0 = 2\pi\cdot
51$\,MHz, and $\Delta \omega_0 \simeq - 2 \pi \cdot 150$\,kHz from
Fig.~4d at the driving voltage $V_g^{ac} = 1.1$\,mV. We then get
$\beta = 4.1 \cdot 10^{24}$\,m$^{-1}$s$^{-2}$ from
Eq.~(\ref{eq:alpha_eff}) by assuming that $\gamma$ is negligible.
A nonzero value of $\gamma >0$ would slightly increase the
estimate for $\beta$.

\section{The dependence of the vibration frequency on  DC gate voltage}

Here, we show that the quadratic nonlinearity of
the restoring force also leads to a shift of the nanoresonator
frequency $\Delta\omega_0$ due to a dc gate voltage. We
start with equation of motion
\begin{equation}
m \partial_t^2 z + 2m\Gamma \partial_t z +m\omega_0^2 z=-m\beta
z^2-m\gamma z^3+\frac{1}{2}\partial_z
C_{g}(V_g^{dc})^2+\frac{1}{2}\partial_z^2
C_{g}(V_g^{dc})^2z\label{eq:equation-of-motion}
\end{equation}
where we assume that the displacement $z$ is
small. Taking $z$ as the sum of a static contribution and an
oscillating contribution, we get in first order in $(V_g^{dc})^2$
\begin{equation}
\Delta\omega_0= \frac{1}{2m\omega_0}[\partial_z
C_{g}\frac{\beta}{\omega_0^2}-\frac{1}{2}\partial_z^2
C_{g}](V_g^{dc})^2.\label{eq:variation-frequence}
\end{equation}
The first term in the bracket, which depends on
the symmetry breaking strength $\beta$, leads to the increase of
$\omega_0$ with $V_g^{dc}$ ($\beta$ is usually positive). This is
the behavior observed for a large majority of nanotube resonators.
By contrast, the second term leads to the decrease of $\omega_0$
with $V_g^{dc}$. This is observed occasionally and is attributed
to nanotubes with a large built-in tension.

Interestingly, the bracket of Eq.~(\ref{eq:variation-frequence})
is the same as that of Eq.~(\ref{eq:major_contributions}) where
the two terms correspond to the low-frequency currents induced by
symmetry breaking and by the nonlinear capacitive coupling,
respectively. In the resonator discussed in the main text,
$\omega_0$ increases with $V_g^{dc}$. This further supports our
finding that the peak in $I_{LF}$ is attributed to symmetry
breaking.

We obtain from Fig.~4b in the main text that the prefactor $a$ in
the relation $\Delta\omega_0 = a(V_g^{dc})^2$ ranges from $4\cdot
10^7$ to $8\cdot 10^7$; we offset $V_g^{dc}$ so that $V_g^{dc}=0$
when $\omega_0$ is minimum. From the length of the nanotube, we
estimate that the effective mass is $\simeq 4$~ag. Neglecting the
second term in Eq.~(\ref{eq:variation-frequence}), we obtain that
$\beta \simeq 3 \pm 1 \cdot 10^{24}$\,m$^{-1}$s$^{-2}$. This value
is comparable to the values estimated above with different
methods.

\begin{figure}
\includegraphics[width=125mm]{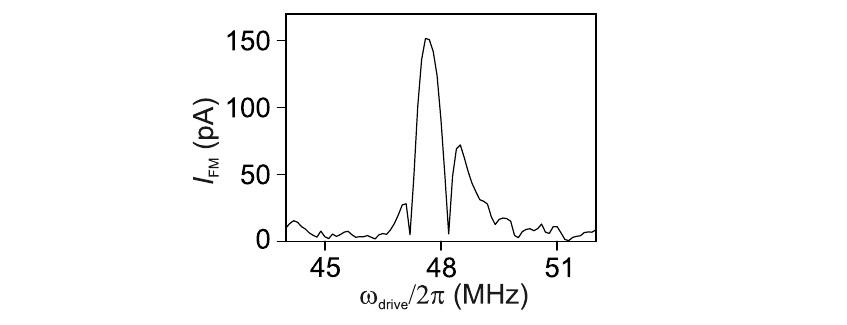}% size for 2column
\caption{\label{Figure S2} \textbf{Mechanical resonance at room
temperature.} Mechanical vibration of the same nanotube device as
in the main text measured at 250~K with the FM technique. The
resonance width $\omega_0 / 2\pi Q$ can be conveniently read out
from the separation between the two minima that are flanking the
main peak~\cite{Gouttenoire2010}. We obtain $Q \sim 50$. $V_g^{dc}
= -0.6$\,V and $V_g^{ac} = 5$\,mV.}
\end{figure}

\section{Fluctuation-induced spectral broadening}

Thermally-induced spectral broadening can be understood from
Eq.~(\ref{eq:equation-of-motion}) if one incorporates into the
right-hand side of this equation a random force $f_T(t)$ that
describes thermal noise. This noise comes from the same coupling
to a thermal reservoir that leads to the friction force $\propto
\Gamma\partial_tz$. From the fluctuation-dissipation relation the
noise is $\delta$-correlated, with $\langle f_T(t)f_T(t')\rangle =
4m\Gamma k_BT\delta(t-t')$.

To gain a qualitative insight into the broadening we assume that
the resonator vibrates as $z_{vibra}\cos(\omega t+\phi)$ with
frequency $\omega$ close to $\omega_0$. We now look at the overall
displacement as $z(t)=z_{vibra}\cos(\omega t+\phi) + \delta z(t)$
and linearize Eq.~(\ref{eq:equation-of-motion}) with respect to
$\delta z(t)$. The left-hand-side will have the same form as for
$z(t)$, i.e., it will describe a resonator with coordinate $\delta
z(t)$ and eigenfrequency  $\omega_0$. In the right-hand-side,
however, there will be a term $-3m\gamma [z_{vibra}\cos(\omega
t+\phi)]^2\delta z(t)$. When averaged over the period
$2\pi/\omega$, this term leads to the shift of the vibration
frequency for $\delta z(t)$ of the form $\omega_0\to
\omega_0+3\gamma z_{vibra}^2/4\omega_0$. This is the well-known
frequency shift of a nonlinear oscillator with vibration
amplitude; a systematic treatment (supplementary reference \cite{Landau_mech}) shows that,
if $z_{vibra}$ is the amplitude of eigenvibrations, to the lowest
order in $z_{vibra}^2$ the frequency shift is $3\gamma
z_{vibra}^2/8\omega_0$, cf. Eq.~(\ref{eq:backbone}).

We now note that the vibrations $z_{vibra}(\cos\omega t+\phi)$ are in fact eigenvibrations induced by noise. They have random phase $\phi$ and also random amplitude. The distribution of this amplitude is of the Boltzmann form, $\propto \exp(-m\omega_0^2z_{vibra}^2/2k_BT)$ for weak resonator nonlinearity. The spread of the vibration amplitudes leads to the effective spread of the vibration eigenfrequencies, with typical width
\begin{equation}
\label{eq:frequency_spread}
\overline{\delta\omega}= 3\gamma_{eff} k_BT/4m\omega_0^3;
\end{equation}
we have replaced here $\gamma$ with $\gamma_{eff}$ to allow for the renormalization of $\gamma$ by the quadratic-nonlinearity term, see Eq.~(\ref{eq:alpha_eff}).

Figure~6 shows the resonance line-shape measured at 250~K. From
the mechanical bandwidth, measured between the two minima that are
flanking the main peak, the apparent quality factor is $\simeq
50$. We change the driving force by a factor up to $4$ and we do
not observe a variation of the bandwidth. Using $\beta= 4.3 \cdot
10^{24}$\,m$^{-1}$s$^{-2}$ and Eq.~(\ref{eq:alpha_eff}), we get an
apparent quality factor of 67 from
Eq.~(\ref{eq:frequency_spread}).

The spread of the eigenfrequencies (\ref{eq:frequency_spread})
leads to a broadening of the resonator spectrum. We emphasize that
this broadening is not related to the vibration decay, it is a
result of the interplay of the resonator nonlinearity and
fluctuations. Moreover, since the distribution of the squared
vibration amplitude, and thus of the vibration eigenfrequency, is
exponential, the spectrum is asymmetric. The overall spectrum in
the presence of nonlinearity and fluctuations, on the one hand,
and decay, on the other hand, is determined by the ratio of
$\overline{\delta\omega}$ and the decay-induced broadening
$\Gamma$. It can be obtained in an explicit form for an arbitrary
$\overline{\delta\omega}/\Gamma$ \cite{Dykman1971}. The frequency spread
of the type (\ref{eq:frequency_spread}) can come also from the
nonlinear coupling of the considered mode to other modes of the
resonator \cite{Dykman1971}. In the context of carbon nanotubes, this
latter mechanism has recently attracted significant attention
\cite{Barnard2012}.

Whereas the internal nonlinearity of the resonator leads to the
change of the shape of the spectrum with increasing temperature,
this is not the case for the nonlinearity associated to the
quadratic dependence of the capacitance on the resonator
displacement (second term in in Eq.~(\ref{eq:nonlinear_energy})).

\begin{figure}
\includegraphics[width=125mm]{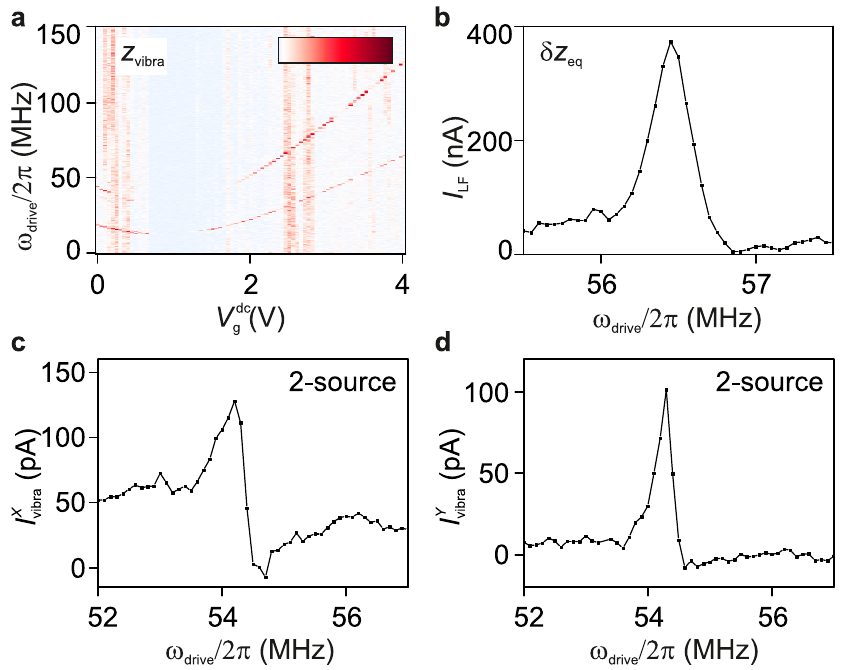}% size for 2column
\caption{\label{Figure S3} \textbf{Measurements for a second
nanotube resonator.} This device has the same geometrical layout
as the one discussed in the main text. All measurements are
performed at $65$\,K. \textbf{a}, Current as a function of
$\omega_{drive}$ and $V_g^{dc}$ measured with the FM technique.
$V_g^{ac} = 1.1$\,mV. Color bar: $0$ (white) to $20$\,pA (dark
red). \textbf{b}, $I_{LF}$ versus $\omega_{drive}$ with $V_g^{ac}
= 0.53$\,mV, $V_{sd}^{dc} = 10$\,mV, and $V_g^{dc} = 2.1$\,V.
\textbf{c}, $X$ quadrature and \textbf{d}, $Y$ quadrature of the
current measured with the 2-source mixing technique.
$I_{vibra}^{X}$ consists of a current proportional to the real
part of the vibrational amplitude in addition to a purely
electrical background current. $I_{vibra}^{Y}$ is proportional to
the imaginary part of the vibrational displacement. $V_g^{ac} =
1.8$\,mV, $V_{sd}^{ac} = 0.6$\,mV, and $V_g^{dc} = 2.1$\,V. A
small shift of the resonance frequency occurred relative to the
measurement in \textbf{b}.}
\end{figure}

\section{Additional device}

In this section, we present data from a second nanotube device. We
verify with atomic force microscopy that the trench width and
depth are the same as for the device presented in the main text
($1.8$\,$\mu$m and $350$\,nm, respectively). The roughness of the
metal electrodes do not allow a measurement of the nanotube
diameter. We therefore use the same estimates for the capacitance
and the mass as for the first device ($C_g = 12$\,aF, $\partial_z
C_g = 5.6$\,pF/m, $\partial^2_z C_g = 21$\,$\mu$F/m$^2$, and $m
\simeq 4$~ag).

In the studied frequency range, we detect two mechanical
resonances that change with an applied gate voltage (Fig.~7a). In
the following, we concentrate on the second visible mode with
$V_g^{dc} = 2.1$\,V. We measure $I_{LF}$ and find a peak at the
resonance frequency (Fig.~7b). In order to determine the
vibration amplitude $z_{vibra}$, we also measure $I_{vibra}^{X}$
and $I_{vibra}^{Y}$ with the 2-source mixing technique (Fig.~7c
and d). We obtain $z_{vibra} = 8.9$\,nm for a driving voltage of
$1.8$\,mV. Unfortunately, the signal-to-noise ratio of the
2-source mixing technique was not sufficient to measure
$z_{vibra}$ directly for the lower driving voltage used to measure
$I_{LF}$. In order to compare the two sets of data, we assume
$z_{vibra} \propto V_g^{ac}$, thus obtaining the scaled value
$z_{vibra} = 2.6$\,nm for $V_g^{ac} = 0.53$\,mV.

We perform the same analysis as for the main device to identify
the origin of $I_{LF}$. We estimate the currents due to the
nonlinearities in the capacitance and in the electrical
conductance, finding $I_{LF}^{capa} = 4.8$\,pA and $I_{LF}^{cond}
= 0.21$\,pA. Both values are far below the measured $I_{LF}^0 =
374$\,pA. In addition to $C_g$, $\partial_z C_g$, and
$\partial^2_z C_g$ mentioned above, we use here $V_{sd}^{dc} =
10$\,mV, $V_{sd}^{ac} = 0.6$\,mV, and $\partial_{V_g}^2 G =
19$\,$\mu$m/V$^2$. For the analysis, we use an offset for
$V_g^{dc}$ such that $V_g^{dc} = 0$ when $\omega_0$ is lowest. As
for the vibration amplitude, we use a scaled value $I_{vibra}^{Y}
= 30$\,pA to account for the difference in the driving voltages.

We now consider symmetry breaking of the vibrations as the origin
of the peak in $I_{LF}$. Using Eq.~(\ref{eq:equilibrium_current}),
we calculate $\beta = 1.8 \cdot 10^{25}$\,m$^{-1}$s$^{-2}$. This
value can be compared to the one we obtain from the dependence of
$\omega_0$ on $V_g^{dc}$. From Eq.~(\ref{eq:variation-frequence}),
we get $\beta = 5.2 \cdot 10^{24}$\,m$^{-1}$s$^{-2}$. For this
device, we have not studied the dependence of $\omega_0$ on
$V_g^{ac}$.

The fluctuation-induced spectral broadening
$\overline{\delta\omega}$, expected from the two values of
$\beta$, leads to a quality factor at room temperature that lies
between $7.3$ and $88$. While we have not measured the quality
factor at room temperature with this device, we note that such
values are very common for nanotube resonators.


\begin{thebibliography}{99}

% =======================================================================================

% =======================================================================================

\bibitem{Sazonova2004}
Sazonova, V., Yaish, Y., \"Ust\"unel, H., Roundy, D., Arias, T. A. \& McEuen, P. L. Tunable carbon nanotube electromechanical oscillator.
\textit{Nature} \textbf{431}, 284-287 (2004).

% =======================================================================================

\bibitem{Lassagne2009}
Lassagne, B., Tarakanov, Y., Kinaret, J., Garcia-Sanchez, D. \& Bachtold, A. Coupling mechanics to charge transport in carbon nanotube mechanical
resonators. \textit{Science} \textbf{325}, 1107-1110 (2009).

% =======================================================================================

\bibitem{Steele2009}
Steele, G. A., H\"uttel, A. K., Witkamp, B., Poot, M., Meerwaldt, H. B., Kouwenhoven, L. P. \& van der Zant, H. S. J. Strong coupling between single-electron tunneling and nanomechanical motion. \textit{Science} \textbf{325}, 1103-1107 (2009).


% =======================================================================================

\bibitem{Gouttenoire2010}
Gouttenoire, V., Barois, T., Perisanu, S., Leclercq, J.-L., Purcell, S. T., Vincent, P. \& Ayari, A. Digital and FM demodulation of a doubly clamped single-walled carbon-nanotube oscillator: towards a nanotube cell phone.
\textit{Small} \textbf{6}, 1060-1065 (2010).

% =======================================================================================

\bibitem{Ganzhorn2013}
Ganzhorn, M. \& Wernsdorfer, W. Dynamics and dissipation induced by single-electron tunneling in carbon nanotube nanoelectromechanical systems. \textit{Phys. Rev. Lett.} \textbf{108}, 175502 (2012).

% =======================================================================================

\bibitem{Benyamini}
Benyamini, A., Hamo, A., Kusminskiy, S. V., von Oppen, F. \&
Ilani, S. Real-space tailoring of the electron-phonon coupling in
ultra-clean nanotube mechanical resonators. Preprint
arXiv:1304.2779 (2013).

% =======================================================================================

\bibitem{Chaste2011}
Chaste, J., Sledzinska, M., Zdrojek, M., Moser, J. \& Bachtold, A. High-frequency nanotube mechanical resonators. \textit{Appl. Phys. Lett.} \textbf{99}, 213502 (2011).

% =======================================================================================

\bibitem{Laird2012}
Laird, E. A., Pei, F., Tang, W., Steele, G. A. \& Kouwenhoven, L. P. A high quality factor carbon nanotube mechanical resonator at 39 GHz.
\textit{Nano Lett.} \textbf{12}, 193-197 (2012).

% =======================================================================================

\bibitem{Huttel2009}
H\"uttel, A. K., Steele, G. A., Witkamp, B., Poot, M., Kouwenhoven, L. P. \& van der Zant, H. S. J. Carbon nanotubes as ultrahigh quality factor mechanical resonators. \textit{Nano Lett.} \textbf{9}, 2547-2552 (2009).

% =======================================================================================

\bibitem{Eichler2011}
Eichler, A., Moser, J., Chaste, J., Zdrojek, M., Wilson-Rae, I. \& Bachtold, A. Nonlinear damping in mechanical resonators made from
carbon nanotubes and graphene. \textit{Nature Nanotech.} \textbf{6}, 339-342 (2011).

% =======================================================================================

\bibitem{Chiu2008}
Chiu, H.-Y., Hung, P., Postma, H. W. Ch. \& Bockrath, M. Atomic-scale mass sensing using carbon nanotube resonators. \textit{Nano Lett.} \textbf{8}, 4342-4346 (2008).

% =======================================================================================

% \bibitem{Wang2010}
% Wang, Z. et al. Phase transitions of adsorbed atoms on the surface of a carbon
% nanotube. \textit{Science} \textbf{327}, 552-555 (2010).

% =======================================================================================

\bibitem{Chaste2012}
Chaste, J., Eichler, A., Moser, J., Ceballos, G., Rurali, R. \& Bachtold, A. A nanomechanical mass sensor with yoctogram resolution. \textit{Nature Nanotech.} \textbf{7}, 301-304 (2012).

% =======================================================================================

\bibitem{Moser}
Moser, J., G\"uttinger, J., Eichler, A., Esplandiu, M. J., Liu, D.
E., Dykman, M. I. \& Bachtold, A. Ultrasensitive force detection
with a nanotube mechanical resonator. \textit{Nature Nanotech.}
\textbf{8}, 493-496 (2013).

% =======================================================================================

%\bibitem{Postma2005}
%H. W. Ch. Postma, I. Kozinsky, A. Husain, and M. L. Roukes, Appl.
%Phys. Lett. \textbf{86}, 223105 (2005).


% =======================================================================================

%\bibitem{Castellanos2012}
%A. Castellanos-Gomez, H. B. Meerwaldt, W. J. Venstra, H. S. J. van
%der Zant, and Gary A. Steele , Phys. Rev. B \textbf{86}, 041402(R)
%(2012).

% =======================================================================================

%\bibitem{Barnard2012}
%A. W. Barnard, V. Sazonova, A. M. van der Zande, and P. L. McEuen,
%Proc. Natl Acad. Sci. USA \textbf{109}, 19093 (2012).

% =======================================================================================

% \bibitem{Rhoads2008}
% J. Rhoads, S. W. Shaw, and K. L. Turner, J. Dyn. Syst. Meas.
% Control \textbf{132}, 034001 (2010).

% =======================================================================================

% \bibitem{Dykman2012}
% \textit{Fluctuating Nonlinear Oscillators}, edited by M. Dykman
% (Oxford University Press, Oxford, England, 2012).

% =======================================================================================

%\bibitem{Garcia2008}
%D. Garcia-Sanchez, A. M. van der Zande, A. San Paulo, B. Lassagne,
%P. L. McEuen, and A. Bachtold, Nano Lett. \textbf{8}, 1399 (2008).

% =======================================================================================

\bibitem{Krivoglaz1966}
Krivoglaz, M. A. \& Pinkevich, I. P. Concerning one mechanism for absorption of low frequency electromagnetic oscillations by localized states in crystals. \textit{Zh. Eksp. Teor. Fiz.} \textbf{51}, 1151-1161 (1966) [Sov. Phys.-JETP \textbf{24}, 772-779 (1967)].

% =======================================================================================

%\bibitem{Dykman1973}
%M. I. Dykman, Sov. Phys. Solid State \textbf{15}, 735 (1973).

% =======================================================================================

\bibitem{Dykman1991}
Dykman, M. I., Mannella, R., McClintock, P. V. E., Soskin, S. M. \& Stocks, N. G. Zero-frequency spectral peaks of underdamped nonlinear oscillators with asymmetric potentials. \textit{Phys. Rev. A} \textbf{43}, 1701-1708 (1991).

% =======================================================================================

% \bibitem{Kittel}
% C. Kittel, \textsl{Introduction to Solid State Physics} (Wiley,
% New York, 7th edition 1996).


% =======================================================================================

\bibitem{Eichler2011B}
Eichler, A., Chaste, J., Moser, J. \& Bachtold, A. Parametric
Amplification and Self-Oscillation in a Nanotube Mechanical
Resonator. \textit{Nano Lett.} \textbf{11}, 2699 (2011).

% =======================================================================================

\bibitem{Sansa}
Sansa, M., Fernandez-Regulez, M., San Paulo, A. \& Perez-Murano,
F. Electrical transduction in nanomechanical resonators based on
doubly clamped bottom-up silicon nanowires. \textit{Appl. Phys.
Lett.} \textbf{101}, 243115 (2012).

% =======================================================================================

\bibitem{Minot2003}
Minot, E. D., Yaish, Y., Sazonova, V., Park, J.-Y., Brink, M. \& McEuen, P. L. Tuning carbon nanotube band gaps with strain. \textit{Phys. Rev. Lett.} \textbf{90}, 156401 (2003).

% =======================================================================================

\bibitem{Stampfer2006}
Stampfer, C., Jungen, A., Linderman, R., Obergfell, D., Roth, S. \& Hierold, C. Nano-electromechanical displacement sensing based on single-walled carbon nanotubes. \textit{Nano Lett.} \textbf{6}, 1449-1453 (2006).

% =======================================================================================

%\bibitem{Chandrahalim2010}
%H. Chandrahalim, C. I. Roman, and C. Hierold, Proceedings 2010 10th IEEE International Conference on Nanotechnology and Joint Symposium with Nano Korea 2010 KINTEX (IEEE-NANO 2010), 778 (2010).


% =======================================================================================

\bibitem{Eichler2012}
Eichler, A., del {\'A}lamo Ruiz, M., Plaza, J. A. \& Bachtold, A. Strong coupling between mechanical modes in a nanotube resonator. \textit{Phys. Rev. Lett.} \textbf{109}, 025503 (2012).

% =======================================================================================

\bibitem{Dykman1971}
Dykman, M. I. \& Krivoglaz, M. A. Classical theory of nonlinear
oscillators interacting with a medium. \textit{Phys. Stat. Sol.
(b)} \textbf{48}, 497-512 (1971).

% ======================================================================================

\bibitem{Barnard2012}
Barnard, A. W., Sazonova, V., van der Zande, A. M. \& McEuen, P. L. Fluctuation broadening in carbon nanotube resonators. \textit{PNAS} \textbf{109}, 19093 (2012).

% =======================================================================================

\bibitem{Dykman1975}
Dykman, M. I. \& Krivoglaz, M. A. Spectral distribution of
nonlinear oscillators with nonlinear friction due to a medium.
\textit{Phys. Stat. Sol. (b)} \textbf{68}, 111-123 (1975).

% =======================================================================================

\bibitem{Croy2012}
Croy, A., Midtvedt, D., Isacsson, A. \& Kinaret, J. M. Nonlinear damping in graphene resonators. \textit{Phys. Rev. B} \textbf{86}, 235435 (2012).

% =======================================================================================

\bibitem{Zaitsev2012}
Zaitsev, S., Shtempluck, O., Buks, E. \& Gottlieb, O. Nonlinear damping in a micromechanical oscillator. \textit{Nonlinear Dynamics} \textbf{67}, 859-883 (2012).

%=======================================================================================

\bibitem{Lifshitz2008}
Lifshitz, R. \& Cross, M. C. Reviews of Nonlinear Dynamics and
Complexity Vol. 1 (Wiley-VCH, 2008), available at
www.tau.ac.il/~ronlif/pubs/RNDC1-1-2008-preprint.pdf

% =======================================================================================

\bibitem{Bunch2007}
Bunch, J. S., van der Zande, A. M., Verbridge, S. S., Frank, I. W., Tanenbaum, D. M., Parpia, J. M., Craighead, H. G. \& McEuen, P. L. Electromechanical resonators from graphene sheets. \textit{Science} \textbf{315}, 490-493 (2007).

% =======================================================================================

\bibitem{Chen2009}
Chen, C., Rosenblatt, S., Bolotin, K. I., Kalb, W., Kim, P., Kymissis, I., Stormer, H. L., Heinz, T. F. \& Hone, J. Performance of monolayer graphene nanomechanical resonators with electrical readout. \textit{Nature Nanotech.} \textbf{4}, 861-867 (2009).

% =======================================================================================

\bibitem{Singh2010}
Singh, V., Sengupta, S., Solanki, H. S., Dhall, R., Allain, A., Dhara, S., Pant, P. \& Deshmukh, M. M. Probing thermal expansion of graphene and modal dispersion at low-temperature using graphene nanoelectromechanical systems resonators. \textit{Nanotechnology} \textbf{21}, 165204 (2010).

% =======================================================================================

\bibitem{Song2012}
Song, X., Oksanen, M., Sillanp\"a\"a, M. A., Craighead, H. G., Parpia, J. M. \& Hakonen, P. J. Stamp transferred suspended graphene mechanical resonators for radio frequency electrical readout. \textit{Nano Lett.} \textbf{12}, 198-202 (2012).

% =======================================================================================

\bibitem{Reserbat2012}
Reserbat-Plantey, A., Marty, L., Arcizet, O., Bendiab, N. \& Bouchiat, V. A local optical probe for measuring motion and stress in a nanoelectromechanical system. \textit{Nature Nanotech.} \textbf{7}, 151-155 (2012).

% =======================================================================================

\bibitem{Chang2009}
Chang, D. E., Regal, C. A., Papp, S. B., Wilson, D. J., Ye, J.,
Painter, O., Kimble, H. J. \& Zoller, P. Cavity opto-mechanics
using an optically levitated nanosphere. \textit{Proc. Natl. Acad.
Sci. U.S.A.} {\bf 107}, 1005-1010 (2009).

% =======================================================================================

\bibitem{Raizen2011}
Li, T., Kheifets, S. \& Raizen, M. G. Millikelvin cooling of an
optically trapped microsphere in vacuum. \textit{Nature Phys.}
{\bf 7}, 527-530 (2011).

% =======================================================================================

\bibitem{Quidant2012}
Gieseler, J., Deutsch, B., Quidant, R. \& Novotny, L. Subkelvin
parametric feedback cooling of a laser-trapped nanoparticle.
\textit{Phys. Rev. Lett.} {\bf 109}, 103603 (2012).

% =======================================================================================

\bibitem{Yamaguchi2012}
Yamaguchi, H., Okamoto, H. \& Mahboob, I. Coherent control of micro/nanomechanical oscillation using parametric mode mixing. \textit{Appl. Phys. Exp.} \textbf{5}, 014001 (2012).

% =======================================================================================

\bibitem{Antonio2012}
Antonio, D., Zanette, D. H. \& L\'opez, D. Frequency stabilization in nonlinear micromechanical oscillators. \textit{Nature Communications}
\textbf{3}, 806 (2012).

% =======================================================================================

\bibitem{Faust2012}
Faust, T., Rieger, J., Seitner, M. J., Krenn, P., Kotthaus, J. P. \& Weig, E. M. Nonadiabatic dynamics of two strongly coupled nanomechanical resonator modes. \textit{Phys. Rev. Lett.} \textbf{109}, 037205 (2012).

% =======================================================================================

\bibitem{Dykman1990}
Dykman, M. I., Mannella, R., McClintock, P. V. E., Soskin, S. M. \& Stocks, N. G. Noise-induced spectral narrowing in nonlinear oscillators. \textit{Europhys. Lett.} 13, 691 (1990).

% =======================================================================================

\bibitem{Kenig2012}
Kenig, E., Cross, M. C., Lifshitz, R., Karabalin, R. B.,
Villanueva, L. G., Matheny, M. H. \& Roukes, M. L. Passive Phase
Noise Cancellation Scheme. \textit{Phys. Rev. Lett.} \textbf{108},
264102 (2012).

% =======================================================================================

\bibitem{Villanueva2011}
Villanueva, L. G., Karabalin, R. B., Matheny, M. H., Kenig, E., Cross, M. C. \& Roukes, M. L. A nanoscale parametric feedback oscillator. \textit{Nano Lett.} \textbf{11}, 5054-5059 (2011).

% =======================================================================================


\end{thebibliography}

\newpage



\begin{thebibliography}{99}

% =======================================================================================
% =======================================================================================

\bibitem{Zant_book}
Meerwaldt, H. B., Steel, G. A. \& van der Zant, H. S. J. in ``\emph {Fluctuating Nonlinear Oscillators}", ed. by M. Dykman (OUP, Oxford 2012), p.~312.

%=====================================

\bibitem{Moser_book}
Moser, J., Eichler, A., Lassagne, B., Chaste, J., Tarakanov, Y., Kinaret, J., Wlson-Rae, I. \& Bachtold, A. \textit{ibid}., p.341.


%==============================================

%\bibitem{Eichler2011}
%A. Eichler, J. Moser, J. Chaste, M. Zdrojek, I. Wilson-Rae, and A. Bachtold, Nature Nanotechnol. \textbf{6}, 339 (2011).

% =======================================================================================

%\bibitem{Eichler2012}
%A. Eichler, M. del {\'A}lamo Ruiz, J. A. Plaza, and A. Bachtold,
%Phys. Rev. Lett. \textbf{109}, 025503 (2012).

%=====================================================

\bibitem{Landau_mech}
Landau, L. D. \& Lifshitz, E. M. \emph{Mechanics} (Elsevier, Amsterdam 2004).

% =======================================================================================

%\bibitem{Lifshitz2008}
%R. Lifshitz and M. C. Cross, \textsl{Reviews of Nonlinear Dynamics
%and Complexity} \textbf{1} (Wiley-VCH, New York, 2008), available
%at www.tau.ac.il/~ronlif/pubs/RNDC1-1-2008-preprint.pdf.

% =======================================================================================






\end{thebibliography}
\end{document}